\documentclass[12pt]{article}
\usepackage{epsfig}
\renewcommand{\d}{\mathrm{d}}
\newcommand{\e}{\mathrm{e}}
\renewcommand{\i}{\mathrm{i}}
\newcommand{\GeV}{\mathrm{GeV}}
\begin{document}
\title{IR-Renormalon Contributions to the\\
Structure Functions $g_3$ and $g_5$}
\author{Benedikt Lehmann-Dronke \and Andreas Sch\"afer}
\date{{\footnotesize Institut f\"ur Theoretische Physik,
Universit\"at Regensburg,\\
Universit\"atsstr.\ 31, 93040 Regensburg, Germany}\\[0.8cm]
June 2, 1998}
\maketitle
\begin{abstract}
We calculate the leading $1/N_f$ perturbative contributions to the
polarized nonsinglet structure functions $g_3$ and $g_5$ to all orders
in $\alpha_s$. The contributions from the first renormalon pole are
determined. It is a measure for the ambiguity of the perturbative
calculation and is assumed to dominate the power corrections. The
corrections $\Delta g_3$ and $\Delta g_5$ are given as functions of
the Bjorken variable $x$ and turn out to be negligable. The anomalous
dimensions of the leading twist operators are obtained in the
next-to-leading order.
\end{abstract}

\vspace{0.5cm}

It is well known that the perturbation series for moments of twist-2
structure functions is an asymptotic one. This property can be studied
in detail in the $1/N_f$-limit, in which the complete series can be
calculated explicitely. Formally this series is given by an integral
over the positive real axis in the Borel plane. This integral is
ambiguous due to singularities on the integration path, the so called
IR-renormalon poles. The residues of these poles are a measure for the
ambiguity of the perturbative series. The so-called hypothesis
of UV-dominance allows furthermore to interpret this ambiguity as an
estimate for the power corrections. the program just sketched was
already applied to all twist-2 structure functions except
$g_3(x,Q^2)$ and $g_5(x,Q^2)$ \cite{da} -- \cite{ me}. In this
contribution we investigate these remaining two
cases. Good experimental data for power corrections to structure
functions exists so far only for $F_L$. The renormalon prediction fits
this data surprisingly well. Let us note that similar renormalon analyses
have recently been applied to a large range of other QCD observables
\cite{be} -- \cite{za}.

The Borel transformation of a perturbative series
\begin{equation}
R=r_0a+r_1a^2+r_2a^3=\ldots=\sum_{n=0}^{\infty}r_na^{n+1}\;,
\quad a=\alpha_s\cdot 4\pi\label{1}
\end{equation}
is defined as
\begin{equation}
B[R](u)=\sum_{n=0}^{\infty}u^n\frac{r_n}{n!}\;.\label{2}
\end{equation}
$R$ can be reobtained from its Borel
transform by an integration over the positive real axis as
\begin{equation}
R=\int_0^{\infty}\d u\,\e ^{-u/a}B[R](u)\;.\label{3}
\end{equation} 
The coefficients of the original power series can also be obtained
individually by taking the derivatives with respect to $u$
\begin{equation}
r_k=\frac{\d^k}{\d u^k}B[R](u)\Bigg|_{u=0}\;.\label{4}
\end{equation}

$B[R]$ has pole singularities on the real axis, the so-called renormalons
\cite{th2}. The poles on the positive $u$-axis, which are called
IR-renormalons because they can be traced back to low momentum
contribution to the loop integrals, lead to ambiguities in the
retransformation (\ref{3}) because it is unclear wether they have to
be passed above or below. The fact that no unambiguous
retransformation exists reflects the
fact that the perturbative expansions are asymptotic \cite{d, bi}. The
ambiguities are of the order of magnitude
\begin{equation}
\Delta R=\e^{-u/a}
\mbox{Res}\Big(B[R](u)\Big)
\Bigg|_{u=\mbox{\footnotesize pole position}}\;,\label{5}
\end{equation}
and can be interpreted as a measure for generic uncertainties of
perturbative predictions or in other words as an estimation for
corrections beyond leading twist perturbation theory \cite{be}.

In connection with the investigation of renormalons the
NNA-approxi\-mation (naive non-Abelianization) \cite{b} is of particular
interest because in the Borel it leads plane to an effective gluon
propagator of a very simple form allowing a calculation to all
orders in the coupling constant. In the NNA-approximation we start
with a restriction to the leading $1/N_f$-terms ($N_f$: number of quark
flavors), which is the sum of all diagrams with only one exchanged
gluon but an arbitrary number of quark loops. The missing
terms are then approximated by the replacement $N_f\rightarrow
-24\pi^2\beta_0=N_f-33/2$, which corresponds to a restriction to
the leading terms of an expansion in the one loop $\beta$-function of
QCD $\beta_0=\frac{1}{(4\pi)^2}(11-\frac{2}{3}N_f)$.
The resummation of all corresponding diagrams leads to the Borel
transformed effective gluon propagator
\begin{equation}
B[g^2D_{\mu\nu}^{ab}(k)](s)=\delta^{ab}\frac{g_{\mu\nu}
-\frac{k_{\mu}k_{\nu}}{k^2}}{k^2}\Bigg(\frac{\mu^2\e^{-c}}{-k^2}\Bigg)^s
\label{6}
\end{equation}
with the new variable $s:=\beta_0u$ \cite{be2, be3, bal}. $c$ is a
renormalization scheme dependent constant, in the
$\overline{\mbox{MS}}$-scheme $c=-5/3$. The expression
(\ref{6}) differs from the original gluon propagator essentially only by
the power of $k^2$ in the denominator. Consequently a
calculation  of a Borel transform in the NNA-approximation in
all orders of the coupling constant is not more complicated than the
corresponding normal next-to-leading-order calculation.

We now apply the described method to the
structure functions $g_3$ and $g_5$ measurable in polarized deep
inelastic lepton-nucleon scattering. These structure functions are
defined by the following terms in the decomposition of the hadronic
scattering tensor
\begin{equation}
W_{\mu\nu}=-\frac{m_N(p_{\mu}S_{\nu}+S_{\mu}p_{\nu})}{p\cdot q}g_3
+\frac{2\,m_NS\cdot q\,g_{\mu\nu}}{p\cdot q}g_5+\ldots\;.\label{7}
\end{equation}
We adopted the conventions of \cite{a}, a comparison with other
definitions used in the literature is given in \cite{bl}. Since the
contributions to $W_{\mu\nu}$ shown in eq.\ (\ref{7}) are parity
violating they involve weak
interactions. We are looking at the case of pure
$Z$-boson exchange and the interference part of $Z$- and
$\gamma$-exchange. In order to avoid operator mixing we consider the
nonsinglet part, which is obtained by taking the difference between
proton- and neutron-structure functions \cite{gr2, p}. To simplify the
notation we write $g_j:=g_j^p-g_j^n\,,\;j=3,5$\,. Neglecting higher
twist contributions, the moments of the structure
functions have the form
\begin{equation}
g_{j,n}:=\int_0^1\d x\,x^ng_j(x,Q^2)=A_{j,n}C_{j,n}(Q^2)\;,\label{8}
\end{equation}
where $A_{j,n}$ are the matrix elements of the leading twist nonsinglet
operators and $C_{j,n}(Q^2)$ the corresponding Wilson
coefficionts. The Wilson coefficients can be calculated using their
connection with the forward Compton scattering amplitude
\begin{eqnarray}
t_{\mu\nu}&=&-\frac{m(p_{\mu}S_{\nu}+S_{\mu}p_{\nu})}{p\cdot q}\,2\,
\sum_na_{3,n}C_{3,n}(Q^2)\omega^{n+1}\nonumber\\
&&+\frac{2m\,S\cdot q\,g_{\mu\nu}}{p\cdot q}\,2\,
\sum_na_{5,n}C_{5,n}(Q^2)\omega^{n+1}+\ldots\;,\label{9}
\end{eqnarray}
where $t_{\mu\nu}$ and $a_{j,n}$ refer to quark states instead of
nucleon states. Adopting a normalization where the non-vanishing
Wilson coefficients take the form
\begin{equation}
C_{j,n}(Q^2)=1+O(g^2)\label{10}
\end{equation}
the matrix elements of the leading twist operators are
\begin{eqnarray}
a_{3,n}&=&2\,V\!A\;,\;\;\;\;n=0,2,4\ldots\nonumber\\
a_{5,n}&=&V\!A\;,\;\;\;\;\;\;\,n=1,3,5\ldots\label{11}
\end{eqnarray}
with the vector coupling constant $V$ and the axial coupling constant
$A$.
The Borel transformed Wilson coefficients are now obtained by the
calculation of $B[t_{\mu\nu}]$ and comparing the result expanded in
$\omega$ with eq.\ (\ref{9}). In the calculations we have to handle the
matrix $\gamma_5$ in $d\neq 4$ dimensions. We use the t'Hooft-Veltman
scheme  $\gamma_5=\i\gamma^0\gamma^1\gamma^2\gamma^3$,
$\{\gamma_5,\gamma^{\mu}\}=0$ for $\mu=0,1,2,3$ and
$[\gamma_5,\gamma^{\mu}]=0$ otherwise \cite{th}. We get
\begin{eqnarray}
B[C_{3,n}](s)&=&
C_{F}\left(\frac{\mu^{2}}{Q^{2}}\right)^{2-\frac{d}{2}}
\left(\frac{\mu^{2}\e^{-c}}{Q^{2}}\right)^{s}\frac{\Gamma(\frac{d}{2})
\Gamma(\frac{d}{2}-2-s)}{(4\pi)^{d/2}\Gamma(s+1)\Gamma(d-1-s)}\nonumber\\
&&\times\Bigg\{(d-2)\frac{\Gamma(s+n+3-\frac{d}{2})}{n!}\nonumber\\
&&\;\;+\bigg(s+2-\frac{d}{2}\bigg)
\Big((d-4)-2\Big)\frac{\Gamma(s+n+2)\Gamma(s+n+4-\frac{d}{2})}
{n!\Gamma(s+n+4)}\nonumber\\
&&\;\;+\frac{d}{2}(6-d)\frac{n\Gamma(s+n+2)\Gamma(s+n+3-\frac{d}{2})}
{n!\Gamma(s+n+4)}\nonumber\\
&&\;\;+\frac{d}{4}\Big(4(d-4)^{2}-4(d-4)(d-2)+(d-2)^{2}\Big)\nonumber\\
&&\;\;\qquad\times\frac{\Gamma(s+n+2)
\Gamma(s+n+3-\frac{d}{2})}{n!\Gamma(s+n+4)}\nonumber\\
&&\;\;+(d-2-s)(d-4)\frac{n\Gamma(s+n+1)\Gamma(s+n+3-\frac{d}{2})}
{n!\Gamma(s+n+3)}\nonumber\\
&&\;\;+4\bigg(s+2-\frac{d}{2}\bigg)\sum_{k=0}^{n}
\frac{\Gamma(s+k+3-\frac{d}{2})}{k!(s+k+1)}\nonumber\\
&&\;\;-2d\sum_{k=0}^{n}
\frac{\Gamma(s+k+3-\frac{d}{2})}{k!(s+k+2)}\nonumber\\
&&\;\;+\bigg((2s+4-d)\frac{d-4}{2}-d\bigg)\sum_{k=0}^{n}
\frac{k\Gamma(s+k+2-\frac{d}{2})}{k!(s+k+1)}\nonumber\\
&&\;\;+d\sum_{k=0}^{n}\frac{k\Gamma(s+k+2-\frac{d}{2})}{k!(s+k+2)}\nonumber\\
&&\;\;+\bigg(\frac{d}{2}-2-s\bigg)
\Big(2(d-4)^{2}-2(d-4)(d-2)+(d-2)^{2}\Big)\nonumber\\
&&\;\;\qquad\times\sum_{k=0}^{n}
\frac{\Gamma(s+k+2-\frac{d}{2})}{k!(s+k+1)}\nonumber\\
&&\;\;+(d\!-\!4\!-\!2s)(d\!-\!2\!-\!s)\frac{d\!-\!4}{d\!-\!2}\sum_{k=0}^{n}
\frac{k\Gamma(s+k+2-\frac{d}{2})}{k!(s+k)}\Bigg\}\label{12}
\end{eqnarray}
for $n=0,2,4\ldots$ and
\begin{eqnarray}
B[C_{5,n}](s)&=&
C_{F}\left(\frac{\mu^{2}}{Q^{2}}\right)^{2-\frac{d}{2}}
\left(\frac{\mu^{2}\e^{-c}}{Q^{2}}\right)^{s}\frac{\Gamma(\frac{d}{2})
\Gamma(\frac{d}{2}-2-s)}{(4\pi)^{d/2}\Gamma(s+1)\Gamma(d-1-s)}\nonumber\\
&&\times\Bigg\{(d-2)\frac{\Gamma(s+n+3-\frac{d}{2})}{n!}\nonumber\\
&&\;\;+\bigg(s+2-\frac{d}{2}\bigg)\Big(2-(d-4)\Big)\frac{\Gamma(s+n+1)
\Gamma(s+n+4-\frac{d}{2})}{n!\Gamma(s+n+3)}\nonumber\\
&&\;\;+\bigg(s+2-\frac{d}{2}\bigg)\Big((d-4)-2\Big)\frac{n\Gamma(s+n+1)
\Gamma(s+n+3-\frac{d}{2})}{n!\Gamma(s+n+3)}\nonumber\\
&&\;\;+\bigg(2(d\!-\!2\!-\!s)(d-4)^{2}
+\frac{1}{2}(6s\!+\!12\!-\!5d)(d-4)(d-2)\nonumber\\
&&\;\;\qquad\qquad\qquad\qquad\qquad\qquad\qquad
+\frac{1}{4}(3d\!-\!8\!-\!4s)(d-2)^{2}\Big)\nonumber\\
&&\;\;\qquad\times\frac{\Gamma(s+n+1)\Gamma(s+n+3-\frac{d}{2})}
{n!\Gamma(s+n+3)}\nonumber\\
&&\;\;+4\bigg(s+2-\frac{d}{2}\bigg)\sum_{k=0}^{n}
\frac{\Gamma(s+k+3-\frac{d}{2})}{k!(s+k+1)}\nonumber\\
&&\;\;-2d\sum_{k=0}^{n}\frac{k\Gamma(s+k+2-\frac{d}{2})}{k!(s+k+1)}\nonumber\\
&&\;\;+\bigg(\frac{d}{2}-2-s\bigg)
\Big(2(d-4)^{2}-2(d-4)(d-2)+(d-2)^{2}\bigg)\nonumber\\
&&\;\;\qquad\times\sum_{k=0}^{n}\frac{\Gamma(s+k+2-\frac{d}{2})}
{k!(s+k+1)}\Bigg\}\label{13}
\end{eqnarray}
for $n=1,3,5\ldots$\,\,.

Since the NNA-approximation is exact in one loop order we get
the next-to-leading-order result from eqs.\ (\ref{12}) and (\ref{13})
by taking $s=0$ according to eq.\ (\ref{4}). An expansion in
$\epsilon=2-\frac{d}{2}$ leads to
\begin{eqnarray}
C_{3,n}&=&1+C_F\frac{g^2}{(4\pi)^2}\Bigg\{\bigg(\frac{1}{\epsilon}
-\gamma+\ln\frac{4\pi Q^2}{\mu^2}\bigg)\nonumber\\
&&\qquad\qquad\qquad\qquad\qquad\times
\bigg(-4+\frac{4}{n+1}+\frac{4}{n+2}-\frac{4}{n+3}+4\,S_n\bigg)\nonumber\\
&&\qquad\qquad\qquad-\frac{3}{2}+\frac{9}{n+1}-\frac{6}{n+3}
+\bigg(3+\frac{2}{n+2}-\frac{4}{n+3}\bigg)S_n\nonumber\\
&&\qquad\qquad\qquad+4\sum_{k=1}^n\frac{1}{k+2}S_k
+2\sum_{k=1}^n\frac{1}{(k+1)(k+2)}S_{k-1}\Bigg\}\;,\label{14}\\
C_{5,n}&=&1+C_F\frac{g^2}{(4\pi)^2}\Bigg\{\bigg(\frac{1}{\epsilon}
-\gamma+\ln\frac{4\pi Q^2}{\mu^2}\bigg)\bigg(-3+\frac{2}{n+1}
+\frac{2}{n+2}+4\,S_n\bigg)\nonumber\\
&&\qquad\qquad\qquad-1-\frac{4}{n+1}+\frac{6}{n+2}
+\bigg(3-\frac{2}{n+1}+\frac{2}{n+2}\bigg)S_n\nonumber\\
&&\qquad\qquad\qquad+8\sum_{k=1}^n\frac{1}{k+1}S_{k-1}\Bigg\}\;,\label{15}
\end{eqnarray}
where $S_n$ is
defined by $S_n:=\sum_{k=1}^n\frac{1}{k}\,$. From the last two equations
we read of the renormalization constants for the corresponding
composite operators (defined by $\mathcal{O}_r=Z^{-1}\mathcal{O}_0$)
in the $\overline{\mbox{MS}}$-scheme \cite{ba}.
\begin{eqnarray}
Z_{g_3,n}&=&1+C_F\frac{g^2}{(4\pi)^2}\bigg(\frac{1}{\epsilon}
-\gamma+\ln 4\pi\bigg)\nonumber\\
&&\qquad\qquad\qquad\times
\bigg(-4+\frac{4}{n+1}+\frac{4}{n+2}-\frac{4}{n+3}+4\,S_n\bigg)\;,\label{16}\\
Z_{g_5,n}&=&1+C_F\frac{g^2}{(4\pi)^2}\bigg(\frac{1}{\epsilon}
-\gamma+\ln 4\pi\bigg)\nonumber\\
&&\qquad\qquad\qquad\times
\bigg(-3+\frac{2}{n+1}+\frac{2}{n+2}+4\,S_n\bigg)\;.\label{17}
\end{eqnarray}
Finally we get for the anomalous dimensions
$\gamma:=\frac{\mu}{Z}\frac{\partial Z}{\partial\mu}\Big|_{g_0}$
(see e.\,g.\ \cite{m}) in one loop order
\begin{eqnarray}
\gamma_{g_3,n}&=&C_F\frac{g^2}{(4\pi)^2}\bigg(8-\frac{8}{n+1}-\frac{8}{n+2}
+\frac{8}{n+3}-8\sum_{k=1}^n\frac{1}{k}\bigg)\;,\label{18}\\
\gamma_{g_5,n}&=&C_F\frac{g^2}{(4\pi)^2}\bigg(6-\frac{4}{n+1}-\frac{4}{n+2}
-8\sum_{k=1}^n\frac{1}{k}\bigg)\;.\label{19}
\end{eqnarray}
To our knowledge these anomalous dimensions have not been calculated
before. Higher order results could be obtained in the---no longer
exact---NNA-approxi\-mation as well using eq.\ (\ref{4}).

To investigate the renormalons we can set $d=4$. From eqs.\ (\ref{12})
and (\ref{13}) we get
\begin{eqnarray}
B[C_{3,n}](s)&=&
C_{F}\left(\frac{\mu^{2}e^{-c}}{Q^{2}}\right)^{s}\frac{1}
{(4\pi)^{2}\Gamma(s+1)(2-s)s}\cdot\frac{1}{s-1}\nonumber\\
&&\times\Bigg\{2\frac{\Gamma(s+n+1)}{n!}
-2s\frac{\Big(\Gamma(s+n+2)\Big)^{2}}{n!\Gamma(s+n+4)}\nonumber\\
&&\;\;+4\frac{n\Gamma(s+n+2)\Gamma(s+n+1)}{n!\Gamma(s+n+4)}
+4\frac{\Gamma(s+n+2)\Gamma(s+n+1)}{n!\Gamma(s+n+4)}\nonumber\\
&&\;\;+2\sum_{k=0}^{n}\Bigg[2s\frac{\Gamma(s+k+1)}{k!(s+k+1)}
-4\frac{\Gamma(s+k+1)}{k!(s+k+2)}-2\frac{k\Gamma(s+k)}
{k!(s+k+1)}\nonumber\\
&&\;\;\;\;\;\;\;\;+2\frac{k\Gamma(s+k)}{k!(s+k+2)}
-2s\frac{\Gamma(s+k)}{k!(s+k+1)}\Bigg]\Bigg\}\;,\label{20}\\
B[C_{5,n}](s)&=&
C_{F}\left(\frac{\mu^{2}e^{-c}}{Q^{2}}\right)^{s}\frac{1}
{(4\pi)^{2}\Gamma(s+1)(2-s)s}\cdot\frac{1}{s-1}\nonumber\\
&&\times\Bigg\{2\frac{\Gamma(s+n+1)}{n!}
+2s\frac{\Gamma(s+n+1)}{n!(s+n+2)}\nonumber\\
&&\;\;-2s\frac{n\Big(\Gamma(s+n+1)\Big)^{2}}{n!\Gamma(s+n+3)}
+4(1-s)\frac{\Big(\Gamma(s+n+1)\Big)^{2}}{n!\Gamma(s+n+3)}\nonumber\\
&&\;\;+4\!\sum_{k=0}^{n}\Bigg[s\frac{\Gamma(s+k+1)}{k!(s\!+\!k\!+\!1)}
\!-\!2\frac{k\Gamma(s+k)}{k!(s\!+\!k\!+\!1)}\!-\!s\frac{\Gamma(s+k)}
{k!(s\!+\!k\!+\!1)}\Bigg]\Bigg\}.\label{21}
\end{eqnarray}
The pole at $s=0$ corresponds to the usual $1/\epsilon$ pole in
dimensional regularization. In both
cases we find two IR-renormalons for $s=1$ and $s=2$. The corrections
corresponding to these renormalons are suppressed by factors
$\frac{1}{Q^2}$ or $(\frac{1}{Q^2})^2$ respectively, which leads to
the hypothesis that they should dominate these power corrections
\cite{s, me, be}. For the dominant pole at
$s=1$ and taking
$\mu^2=Q^2$ we find for the residues
\begin{eqnarray}
\mbox{Res}\Big(B[C_{3,n}](s)\Big)\Bigg|_{s=1} &=&
-\frac{C_{F}e^{-c}}{(4\pi)^{2}}
\Bigg\{2n+6-\frac{4}{n+1}-\frac{4}{n+2}\nonumber\\
&&\;\;\;\;\;\;\;\;-\frac{16}{n+3}+\frac{24}{n+4}
-4\sum_{k=1}^{n}\frac{1}{k}\Bigg\}\;,\label{22}\\
\mbox{Res}\Big(B[C_{5,n}](s)\Big)\Bigg|_{s=1} &=&
-\frac{C_{F}e^{-c}}{(4\pi)^{2}}
\Bigg\{2n+10-\frac{8}{n+1}\nonumber\\
&&\;\;\;\;\;\;\;\;-\frac{4}{n+2}-\frac{8}{n+3}
-8\sum_{k=1}^{n}\frac{1}{k}\Bigg\}\;,\label{23}
\end{eqnarray}
which are connected with the renormalon contributions according to
eq.\ (\ref{5}) by
\begin{equation}
\Delta C_{j,n}(Q^2)=\pm\Bigg(\frac{\Lambda^2}{Q^2}\Bigg)\frac{1}{\beta_0}
\mbox{Res}\Big(B[C_{j,n}](s)\Big)\Bigg|_{s=1}
+O\Bigg(\frac{g^2}{Q^2},\frac{1}{Q^4}\Bigg)\;.\label{24}
\end{equation}
The sign of the corrections remains unknown. Correspondingly we get
according to eq.\ (\ref{8}) for the  complete structure functions
\begin{equation}
g_{j,n}(Q^2)=A_{j,n}\Big[\sum_{k=0}^{N_0}C_{j,n}^k(Q^2)(g^2)^k+
\Delta C_{j,n}(Q^2)\Big]\label{25}
\end{equation}
with the perturbative expansion of the Wilson coefficients $C_{j,n}
\!=\!\sum_k\!C_{j,n}^k(g^2)^k$ and for the renormalon corrections of
the same structure functions
\begin{equation}
\Delta g_{j,n}(Q^2)=A_{j,n}\Delta C_{j,n}(Q^2)\;.\label{26}
\end{equation}
The unknown matrix elements  $A_{j,n}$ are eliminated taking the ratio
\begin{eqnarray}
\frac{\Delta g_{j,n}(Q^{2})}{g_{j,n}(Q^{2})} &=&
\frac{\Delta C_{j,n}(Q^{2})}
{\sum_{k=0}^{N_{0}}C_{j,n}^{k}(g^{2})^{k}
+\Delta C_{j,n}(Q^2)}\nonumber\\
&=&
\frac{\pm\left.\left(\frac{\Lambda^{2}}{Q^{2}}\right)\frac{1}{\beta_{0}}
\mbox{Res}\Big(B[C_{j,n}](s)\Big)\right|_{s=1}
+O\left(\frac{g^{2}}{Q^{2}},\frac{1}{Q^{4}}\right)}
{1+\sum_{k=1}^{N_{0}}C_{j,n}(Q^{2})^{k}(g^{2})^{k}
+\Delta C_{j,n}(Q^{2})}\nonumber\\
&=&
\left[\pm\left.\left(\frac{\Lambda^{2}}{Q^{2}}\right)\frac{1}{\beta_{0}}
\mbox{Res}\Big(B[C_{j,n}](s)\Big)\right|_{s=1}
+O\left(\frac{g^{2}}{Q^{2}},\frac{1}{Q^{4}}\right)\right]\nonumber\\
&&\quad
\times\left[1+O\left(g^{2},\frac{1}{Q^{2}}\right)\right]\nonumber\\
&=&
\pm\left.\left(\frac{\Lambda^{2}}{Q^{2}}\right)\frac{1}{\beta_{0}}
\mbox{Res}\Big(B[C_{j,n}](s)\Big)\right|_{s=1}
+O\left(\frac{g^{2}}{Q^{2}},\frac{1}{Q^{4}}\right)\;.\label{27}
\end{eqnarray}
So in leading order the corrections are given as
\begin{equation}
\Delta g_{j,n}(Q^{2})=\pm\left.\left(\frac{\Lambda^{2}}{Q^{2}}\right)
\frac{1}{\beta_{0}}\mbox{Res}\Big(B[C_{j,n}](s)\Big)\right|_{s=1}
\cdot g_{j,n}(Q^{2})\;.\label{28}
\end{equation}
\begin{figure}
\begin{minipage}[t]{6cm}
\hspace{-2cm}
\epsfig{file=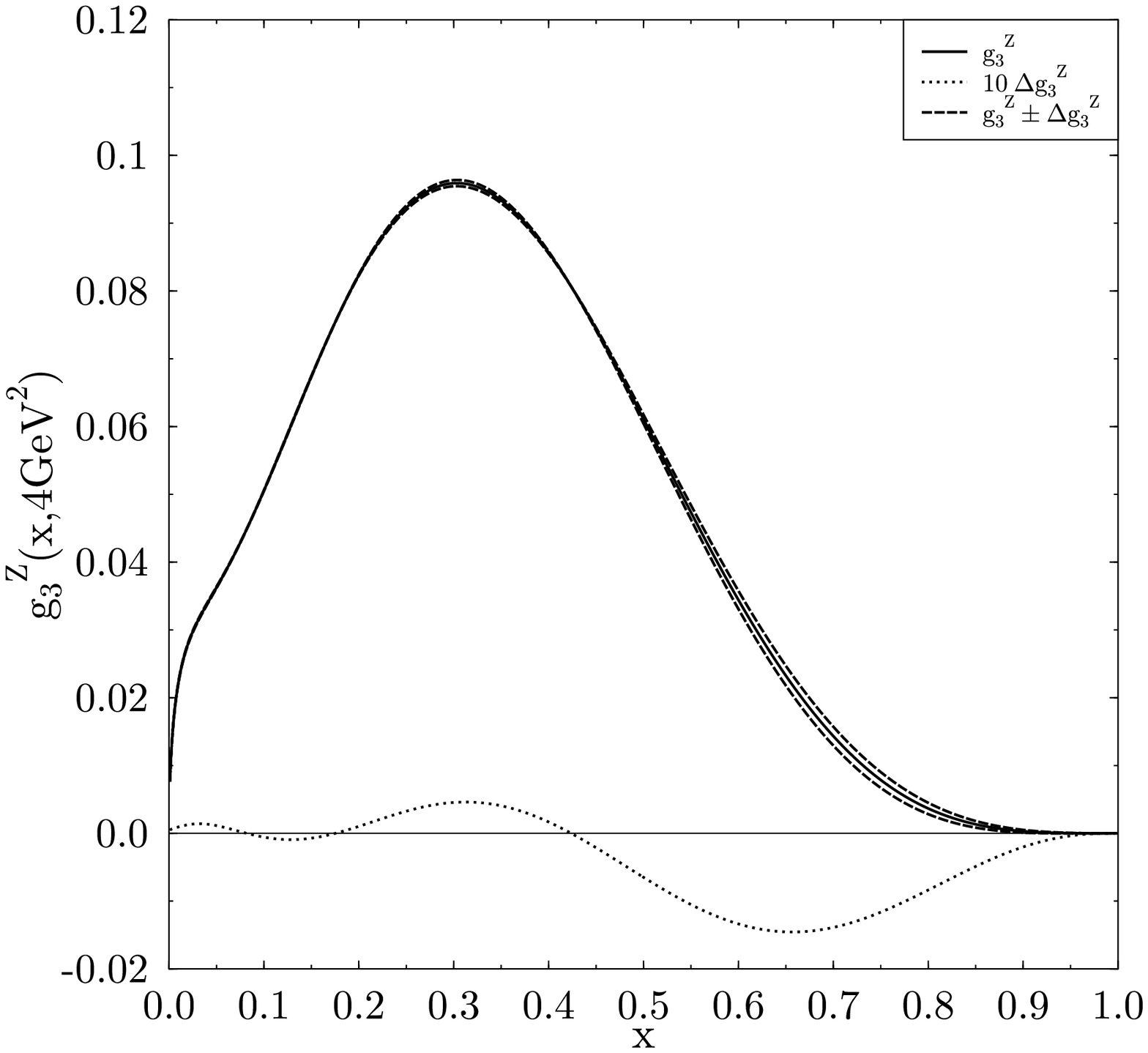, height=6cm}
\caption{The fit for $g_3^Z$ (full line) and the corresponding
renormalon contribution multiplied by a factor $10$ (dotted
line). The dashed lines show the size of the ambiguity for $g_3^Z$,
i.\,e.\ $g_3^Z\pm\Delta g_3^Z$.}
\label{f1}
\end{minipage}
\begin{minipage}[t]{6cm}
\hspace{-2cm}
\epsfig{file=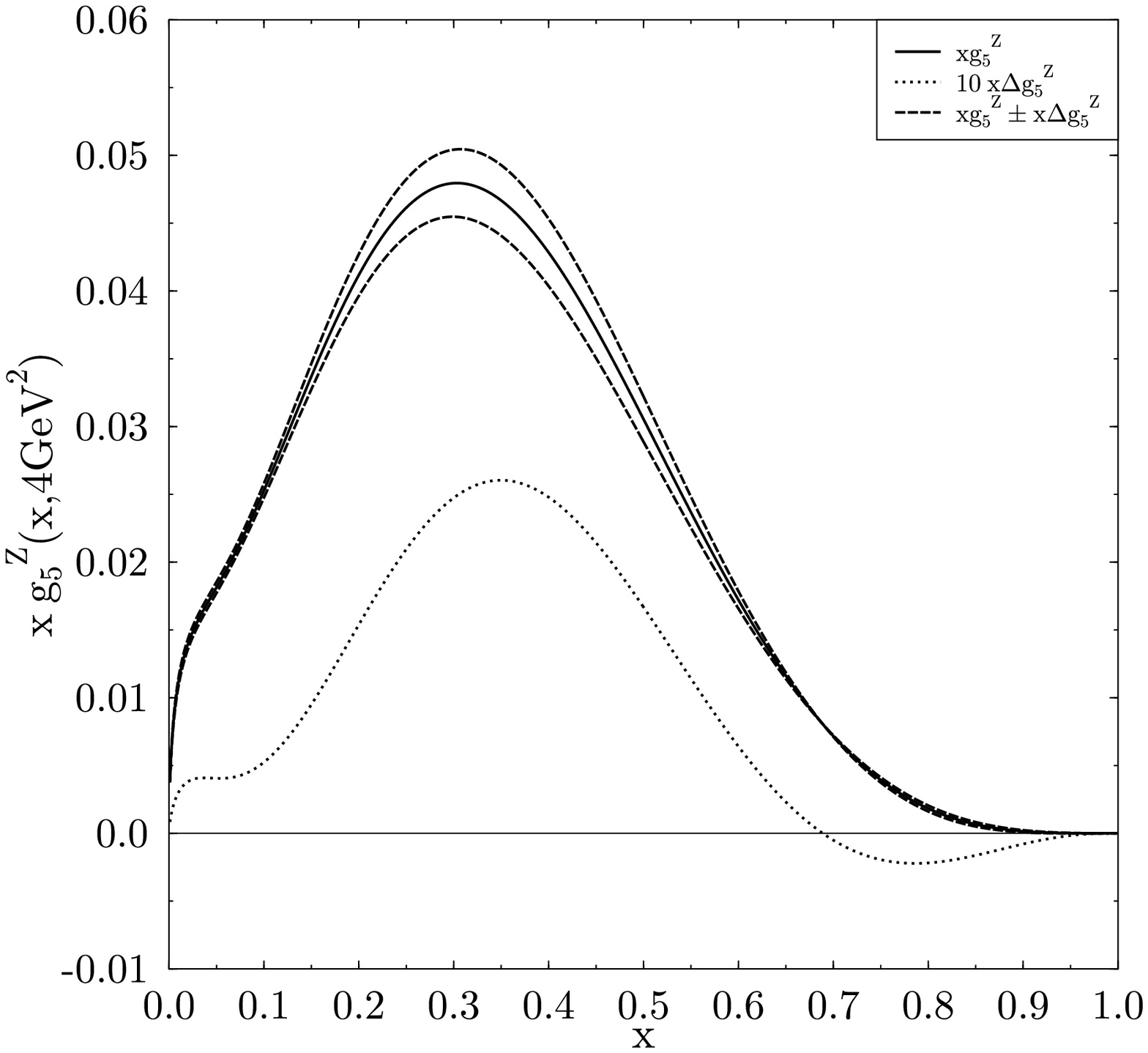, height=6cm}
\caption{The same as fig.\ \ref{f1} for $g_5^Z$.}
\label{f2}
\end{minipage}
\begin{minipage}{6cm}
\hspace{-2cm}
\epsfig{file=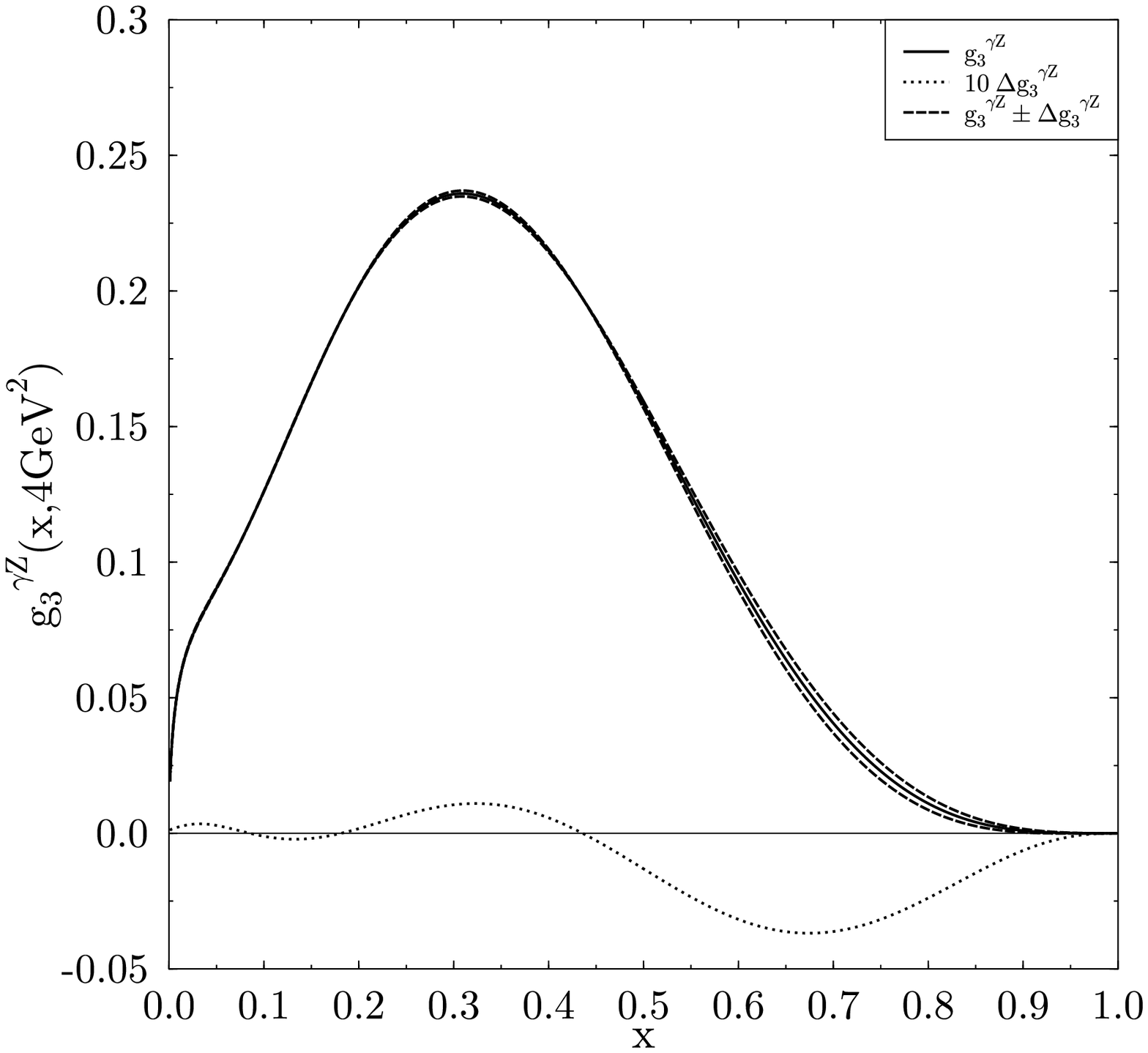, height=6cm}
\caption{The same as fig.\ \ref{f1} for $g_3^{\gamma Z}$.}
\label{f3}
\end{minipage}
\hspace{1.4cm}
\begin{minipage}{6cm}
\hspace{-2cm}
\epsfig{file=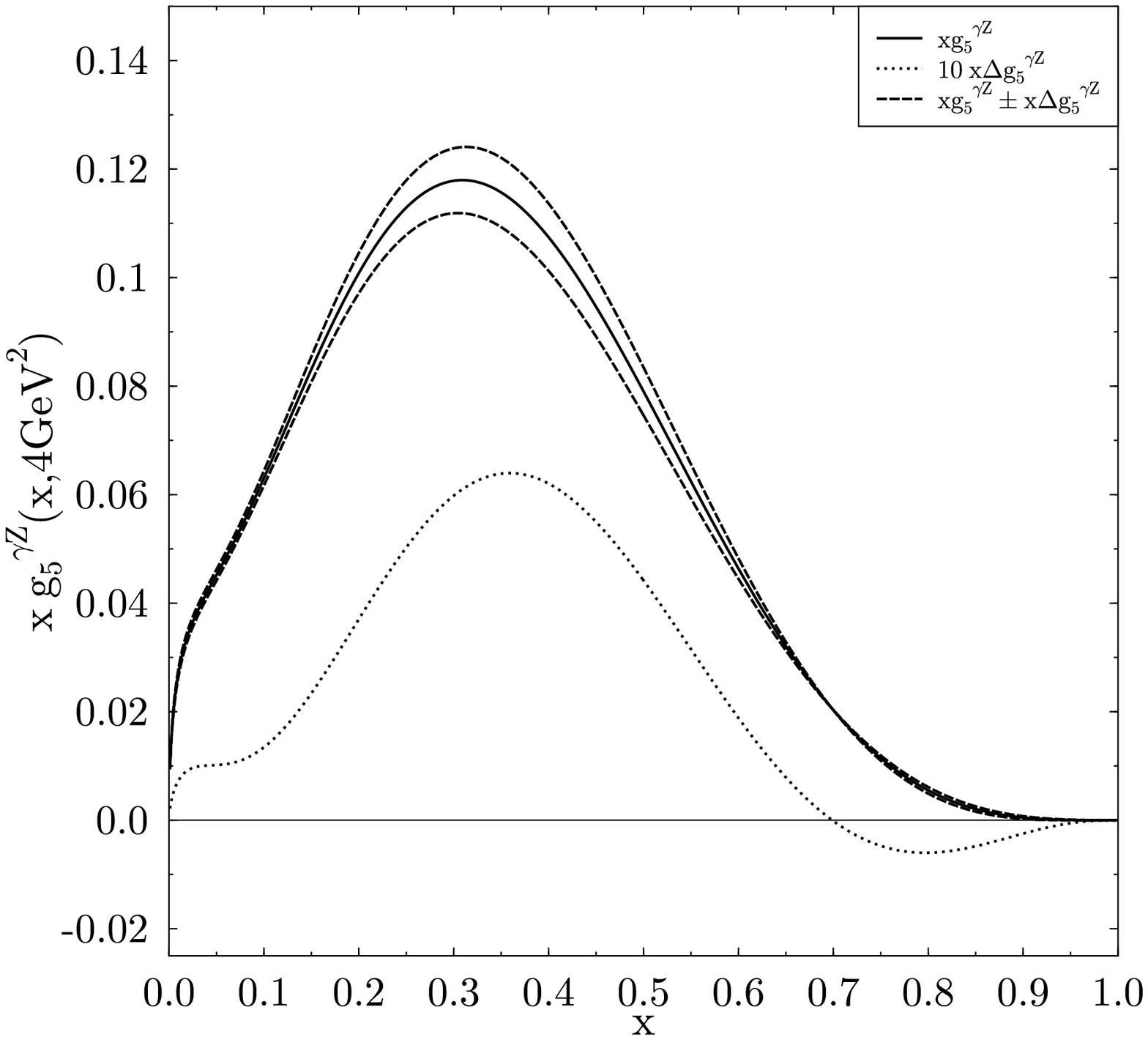, height=6cm}
\caption{The same as fig.\ \ref{f1} for $g_5^{\gamma Z}$.}
\label{f4}
\end{minipage}
\end{figure}
The deternination of all moments is equivalent to expressing $\Delta
g_j(x)$ as a convolution
\begin{equation}
\Delta g_{j}(x)=\pm\left(\frac{\Lambda^{2}}{Q^{2}}\right)\frac{1}{\beta_{0}}
\int_{x}^{1}\frac{\d y}{y}C_{j}^{IR1}(y)g(x/y)\;,\label{29}
\end{equation}
where the functions $C_j^{IR1}(y)$ defined by
\begin{equation}
\mbox{Res}\Big(B[C_{j,n}](s)\Big)\Bigg|_{s=1}
=\int_{0}^{1}\d y\,y^{n}C_{j}^{IR1}(y)\label{30}
\end{equation}
are obtained from eqs.\ (\ref{20}) and (\ref{21}):
\begin{eqnarray}
C_{3}^{IR1}(y)\!=\!\frac{C_{F}\e^{-c}}{(4\pi)^{2}}
\left\{2\delta^{\prime}(y\!-\!1)\!-\!6\delta(y\!-\!1)\!
+\!4\!+\!4y\!+\!16y^{2}\!-\!24y^{3}\!
-\!\frac{4}{(1\!-\!y)_{+}}\right\},\label{31}\\
C_{5}^{IR1}(y)\!=\!\frac{C_{F}\e^{-c}}{(4\pi)^{2}}
\left\{2\delta^{\prime}(y\!-\!1)\!-\!10\delta(y\!-\!1)\!+\!8\!
+\!4y\!+\!8y^{2}\!
-\!\frac{8}{(1\!-\!y)_{+}}\right\},\hspace{1.2cm}\label{32}
\end{eqnarray}
where $\frac{1}{(1-y)_+}$ is defined by
$\int_{0}^{1}\d y\,f(y)\frac{1}{(1-y)_{+}}
=\int_{0}^{1}\d y\frac{f(y)-f(1)}{1-y}$\,.
We use the quark distributions given in \cite{g} and the parton model
expressions
\begin{eqnarray}
g_{3}^{Z} &=& 2x\sum_{q}g_{V}^{q}g_{A}^{q}
(\Delta q-\Delta\overline{q})\;,\label{33}\\
2xg_{5}^{Z} &=& g_{3}^{Z}\;,\label{34}\\
g_{5}^{\gamma Z} &=& 2x\sum_{q}e^{q}g_{A}^{q}
(\Delta q-\Delta\overline{q})\;,\label{35}\\
2xg_{5}^{\gamma Z} &=& g_{5}^{\gamma Z}\:.\label{36}
\end{eqnarray}
We choose the momentum transfer to be $Q^2=4\,\GeV^2$. The
integrals in eq.\ (\ref{29}) are evaluated numerically and the results are
plotted in the figures \ref{f1} to \ref{f4}.

We have thus completed our analysis of the renormalon ambiguities for
all twist-2 structure functions.

{\bf Acknoledgement:} This work was supported by BMBF.

\end{document}